\documentstyle[preprint,eqsecnum,aps]{revtex}
\begin{document}
\tightenlines

\draft
\title{Generalized Reissner--Nordstr\"om solution in Metric--Affine Gravity} 
\author{Alfredo Mac\'{\i}as$^{\diamond}$$^\star$
\thanks{E-mail: amac@xanum.uam.mx,\, macias@nuclecu.unam.mx}
and Jos\'e Socorro$^{\$}$\thanks{E-mail: socorro@ifug4.ugto.mx}\\ 
$^{\diamond}$ Departamento de F\'{\i}sica,\\
Universidad Aut\'onoma Metropolitana--Iztapalapa,\\
Apartado Postal 55-534, 09340 M\'exico, D.F., MEXICO\\
$^\star$ Instituto de Ciencias Nucleares, 
Universidad Nacional Aut\'onoma de M\'exico \\ 
Apartado Postal 70-543, 04510 M\'exico D.F, MEXICO\\
$^{\$}$ Instituto de F\'{\i}sica de la Universidad de Guanajuato,\\
Apartado Postal E-143, C.P. 37150, Le\'on, Guanajuato, MEXICO
}

\date{\today}

\maketitle

\begin{abstract}
We present the generalized Reissner--Nordstr\"om solution of 
the field equations of metric--affine gravity (MAG), endowed with electric 
and magnetic charges, as well as with gravito--electric and gravito--magnetic 
charges and a cosmological constant term. Moreover, the case $M=e_o$, 
i.e. mass equal to electric charge and $\lambda=0$, corresponds to an 
electrically and magnetically charged monopole. Also further multipole 
solutions are obtained.
The charge assignments of the solutions is discussed.
{\bf file mono8.tex 04.04.1999}.

\end{abstract}


\section{Introduction}

Examples in which spacetime might become non--Riemannian above
Planck energies occur in string theory \cite{ft,cfmp,g,gm,gm1}
or in the very early universe in the inflationary 
model \cite{gu,gu1,lin,lin1,ls}. The simplest such geometry is 
metric--affine geometry, in which nonmetricity appears as a field strength, 
side by side with curvature and torsion \cite{nehe}.
The Einstein--Cartan theory yields a non--Riemannian spacetime
already for $10^{52} g/cm^3$, well below the Planck density of about
$10^{92} g/cm^3$ \cite{rmp}.

As it is well known \cite{PR}, MAG represents a gauge theory of the 
4--dimensional affine group enriched by the existence of a metric. 
As a gauge theory, it finds its appropriate form if expressed with 
respect to arbitrary frames or {\em coframes}. We have then the metric 
$g_{\alpha\beta}$, the coframe $\vartheta^\alpha$, and the connection 1--form
$\Gamma_\alpha{}^\beta$ (with values in the Lie algebra of the
4--dimensional linear group $GL(4,R)$) as independent field
variables. Spacetime is described by a metric--affine geometry with
the gravitational {\em field strengths} nonmetricity $Q_{\alpha\beta}
:=-Dg_{\alpha\beta}$, torsion $T^\alpha:=D\vartheta^\alpha$, and
curvature $R_\alpha{}^\beta:= d\Gamma_\alpha{}^\beta
-\Gamma_\alpha{}^\gamma\wedge\Gamma_\gamma{}^\beta$. 

On the other hand, we do not believe that at the
present state of the universe the geometry of spacetime is described
by a metric--affine one. We rather think, and there is good
experimental evidence, that the present--day geometry is
metric--compatible, i.e., its nonmetricity vanishes. In earlier epochs
of the universe, however, when the energies of the cosmic ``fluid''
were much higher than today, we expect scale invariance to prevail ---
and the canonical dilation (or scale) current of matter, the trace of
the hypermomentum current $\Delta^\gamma{}_\gamma$, is coupled,
according to MAG, to the Weyl covector $Q^\gamma{}_\gamma$. By the
same token, shear type excitations of the material multispinors (Regge
trajectory type of constructs) are expected to arise, thereby
liberating the (metric--compatible) Riemann--Cartan spacetime from its
constraint of vanishing nonmetricity $Q_{\alpha\beta}=0$ . Tresguerres
\cite{Tres3} has proposed a simple cosmological model of Friedmann
type which carries a metric--affine geometry at the beginning of the
universe, the nonmetricity of which dies out exponentially in time.
That is the kind of thing we expect.

Moreover, for getting a deeper understanding of the meaning and the possible
consequences of MAG, a search for exact solutions appears
indispensable. The search for exact solutions of MAG is still in its 
infancy. It began with the work of
Tres\-guerres\cite{Tres14,Tres15}, Tucker and Wang\cite{TW}, Obukhov
et al. \cite{heh96}, Vlashinsky et al. \cite{heh961}, and of Puntigam et
al. \cite{PLH97}.  Mac\'{\i}as et al. \cite{mms98}, and Socorro et al.
\cite{slmm} mapped the Einstein--Maxwell sector of dilaton--gravity,
emerging from low energy string theory, and found new soliton and
multipole solutions of MAG.  However, it is important to note that in
order to incorporate the scalar dilaton field, one could, for
instance, generalize the torsion kink of Baekler et al. \cite{kink},
an exact solution with an external massless scalar field, or one could
turn to the axi--dilatonic sector of MAG \cite{dt95}. 
Garc\'{\i}a et al. \cite{pd98} found the Pleba\'nski--Demia\'nski 
class of solutions in MAG.
Moreover,
solutions implying the existence of torsion shock waves have already
been found by Garc\'{\i}a et al. \cite{glmms}.
For a review of all known exact solutions in MAG see Ref.
\cite{hema99}.

It has been demonstrated that for {\em each} Einstein--Maxwell solution 
(metric plus electromagnetic potential 1--form), if the electric charge is 
replaced by the strong gravito--electric charge and if a suitable 
constraint on the coupling constants is postulated, an exact solution of 
MAG can be created. Even more so, the  MAG model considered here can be reduced  
to an {\em effective} Einstein--Proca system \cite{De96,oveh97}. Now the meaning of 
the constraint between the different coupling 
constants becomes immediately clear: If we put $m^2=0$, then the
Einstein--Proca system becomes an Einstein--Maxwell system.
Moreover, it is quite useful to find electrically charged versions of 
a MAG solution \cite{helo,aas} in order to illustrate the coupling of the 
electromagnetic field to the post--Riemannian structures of a 
metric--affine spacetime.  

In this paper, we present a generalized Reissner--Nordstr\"om solution in 
MAG endowed with electric, magnetic, gravito--electric and gravito--magnetic
charges and cosmological constant. Moreover, the electrically and 
magnetically charged gravito--electric and gravito--magnetic MAG 
monopole solutions contained on it are also discussed.
It is worthwhile to stress the well know fact \cite{gama} that in the general 
relativistic context, the Reissner--Nordstr\"om solution is very 
important since it represents the gravitational field of a static charged 
mass. It is the {\em only static} solution which can be thought of as a 
static black hole endowed with mass and electric charge. 
Switching off the charge, it reduces to the Schwarzschild solution.

The plan of the paper is as follows: In Sec. II we review the 
MAG Lagrangian. In Sec. III we present the solitonic monopole field 
configuration in MAG. In Sec. IV electrically and magnetically charged strong 
gravito--magnetic monopole solutions are presented.
In Sec. V we transform the monopole solution to Schwarzschild--like 
coordinates. In Sec. VI a generalized Reissner--Nordstr\"om solution in MAG 
is presented. In Sec. VII we display further multipole solutions.
In Sec. VIII the charge assignment of the solutions is given. 
In Sec. IX the results are discussed.
\section{Quadratic MAG Lagrangian}

Propagating post--Riemannian gauge interactions in MAG can be consistently
constructed by adding terms quadratic in in $Q_{\alpha\beta}$,
$T^\alpha$, $R_\alpha{}^\beta$ to the Hilbert--Einstein type Lagrangian
and the term with the cosmological constant.

In the first order formalism we are using, higher order terms
in the gauge field strengths
$Q_{\alpha\beta}$, $T^\alpha$, $R_\alpha{}^\beta$, i.e.\
cubic and quartic ones etc.\ would preserve the second order of the
field equations. However, the {\em quasilinearity} of the gauge field
equations would be destroyed and, in turn, the Cauchy problem would be
expected to be ill--posed. Therefore we do not go beyond a gauge
Lagrangian which is {\em quadratic} in the gauge field strengths. 
Moreover, a quadratic Lagrangian is already so messy that it would be hard to
handle a still more complex one anyway.

The {\em most general parity conserving quadratic}
Lagrangian which is expressed in terms of the $4+3+11$ irreducible
pieces of $Q_{\alpha\beta}$, $T^\alpha$, $R_\alpha{}^\beta$,
respectively (see \cite{PR}):
\begin{eqnarray} 
\label{lobo} V_{\rm MAG}&=&
\frac{1}{2\kappa}\,\left[-a_0\,R^{\alpha\beta}\wedge\eta_{\alpha\beta}
  -2\lambda\,\eta+T^\alpha\wedge{}^*\!\left(\sum_{I=1}^{3}a_{I}\,^{(I)}
    T_\alpha\right)\right.\nonumber\\ &+&\left.
  2\left(\sum_{I=2}^{4}c_{I}\,^{(I)}Q_{\alpha\beta}\right)
  \wedge\vartheta^\alpha\wedge{}^*\!\, T^\beta + Q_{\alpha\beta}
  \wedge{}^*\!\left(\sum_{I=1}^{4}b_{I}\,^{(I)}Q^{\alpha\beta}\right)\right.
\nonumber \\&+&
b_5\bigg.\left(^{(3)}Q_{\alpha\gamma}\wedge\vartheta^\alpha\right)\wedge
{}^*\!\left(^{(4)}Q^{\beta\gamma}\wedge\vartheta_\beta \right)\bigg]
\nonumber\\&- &\frac{1}{2\rho}\,R^{\alpha\beta} \wedge{}^*\!
\left(\sum_{I=1}^{6}w_{I}\,^{(I)}W_{\alpha\beta}
  +w_7\,\vartheta_\alpha\wedge(e_\gamma\rfloor
  ^{(5)}W^\gamma{}_{\beta} ) \nonumber\right.\\&+& \left.
  \sum_{I=1}^{5}{z}_{I}\,^{(I)}Z_{\alpha\beta}+z_6\,\vartheta_\gamma\wedge
  (e_\alpha\rfloor ^{(2)}Z^\gamma{}_{\beta}
  )+\sum_{I=7}^{9}z_I\,\vartheta_\alpha\wedge(e_\gamma\rfloor
  ^{(I-4)}Z^\gamma{}_{\beta} )\right)
\label{6}\,.  
\end{eqnarray}
The constant $\lambda$ is the cosmological constant, $\rho$ the strong
gravity coupling constant, the constants $ a_0, \ldots a_3$, $b_1,
\ldots b_5$, $c_2, c_3,c_4$, $w_1, \ldots w_7$, $z_1, \ldots z_9$ are
dimensionless. We have introduced in the curvature square term the
irreducible pieces of the antisymmetric part $W_{\alpha\beta}:=
R_{[\alpha\beta]}$ and the symmetric part $Z_{\alpha\beta}:=
R_{(\alpha\beta)}$ of the curvature 2--form.  In $Z_{\alpha\beta}$, we
have the purely {\em post}--Riemannian part of the curvature. Note the
peculiar cross terms with $c_I$ and $b_5$.

The first and the second field equations of MAG read 
\begin{eqnarray}
DH_{\alpha}-
E_{\alpha}&=&\Sigma_{\alpha}\,,\label{first}\\ DH^{\alpha}{}_{\beta}-
E^{\alpha}{}_{\beta}&=&\Delta^{\alpha}{}_{\beta}\,,
\label{second}
\end{eqnarray}
where $\Sigma_{\alpha}$ and $\Delta^{\alpha}{}_{\beta}$ are the
canonical energy--momentum and hypermomentum current three--forms
associated with matter. 
The left hand sides of
(\ref{first})--(\ref{second}) involve the two--forms of the gravitational gauge 
field momenta $H_{\alpha}$ and $H^{\alpha}{}_{\beta}$
(gravitational ``excitations"). We find them, together with $M^{\alpha\beta}$,
by partial differentiation of the Lagrangian (\ref{lobo}):
\begin{eqnarray}
M^{\alpha\beta}&:=&-2{\partial V_{\rm MAG}\over \partial Q_{\alpha\beta}}
= -{2\over\kappa}\left\{{}^*\! \left(\sum_{I=1}^{4}b_{I}{}^{(I)} 
Q^{\alpha\beta}\right)+{1\over2}\,
b_5\left[\vartheta^{(\alpha}\wedge{}^*(Q\wedge\vartheta^{\beta)})
-{1\over 4}\,g^{\alpha\beta}\,^*(3Q+\Lambda)\right]\right. \nonumber\\ 
&+& \left. c_{2}\,\vartheta^{(\alpha}\wedge{}^*{}\!^{(1)}T^{\beta)} +
c_{3}\,\vartheta^{(\alpha}\wedge{}^*{}\!^{(2)}T^{\beta)} + {1\over 4}
(c_{3}-c_{4})\,g^{\alpha\beta}{}^*\!\, T \right\}\, ,\label{M1}\\ 
H_{\alpha}&:=&-{\partial V_{\rm MAG}\over \partial T^{\alpha}} = -
{1\over\kappa}\, {}^*\!\left[\left(\sum_{I=1}^{3}a_{I}{}^{(I)}
T_{\alpha}\right) + \left(\sum_{I=2}^{4}c_{I}{}^{(I)}
Q_{\alpha\beta}\wedge\vartheta^{\beta}\right)\right],\label{Ha1}\\ 
H^{\alpha}{}_{\beta}&:=& - {\partial V_{\rm MAG}\over \partial
R_{\alpha}{}^{\beta}}= {a_0\over 2\kappa}\,\eta^{\alpha}{}_{\beta} 
+ {\cal W}^{\alpha}{}_{\beta} + {\cal Z}^{\alpha}{}_{\beta},\label{Hab1}
\end{eqnarray}
where we introduced the abbreviations
\begin{equation}
{\cal W}_{\alpha\beta}:= {}^*\!
\left(\sum_{I=1}^{6}w_{I}{}^{(I)}W_{\alpha\beta} \right),\quad\quad
{\cal Z}_{\alpha\beta}:= {}^*\!
\left(\sum_{I=1}^{5}z_{I}{}^{(I)}Z_{\alpha\beta} \right).
\end{equation}
The three--forms $E_{\alpha}$ and $E^{\alpha}{}_{\beta}$
describe the canonical energy--mo\-men\-tum and hypermomentum  
currents of the gauge fields themselves. They can be written as follows 
\cite{PR}:
\begin{eqnarray}
E_{\alpha} & = & e_{\alpha}\rfloor V_{\rm MAG} + (e_{\alpha}\rfloor  
T^{\beta})
\wedge H_{\beta} + (e_{\alpha}\rfloor R_{\beta}{}^{\gamma})\wedge
H^{\beta}{}_{\gamma} + {1\over 2}(e_{\alpha}\rfloor Q_{\beta\gamma})
M^{\beta\gamma}\\ 
E^{\alpha}{}_{\beta} & = & - \vartheta^{\alpha}\wedge H_{\beta} - 
M^{\alpha}{}_{\beta}\, ,
\end{eqnarray}
where $e_{\alpha}\rfloor$ denotes the interior product with a form.
\section{Monopole field configuration in MAG}

We search for exact solutions of the field equations belonging to
the Lagrangian
\begin{equation}\label{Ltot}
L=V_{\rm MAG}+V_{\rm Max}\,,
\end{equation}
where $ V_{\rm Max}=-(1/2)F\wedge\hspace{-0.8em}  {\phantom{F}}^{\star}F$
is the Lagrangian of the Maxwell field and $F=dA$ is the electromagnetic 
field strength.
Since the matter part $V_{\rm Max}$ does not depend
on the connection $\Gamma_\alpha{}^\beta$, the hypermomentum
$\Delta^\alpha{}_\beta:=\delta V_{\rm Max}/\delta
\Gamma_\alpha{}^\beta$ vanishes, i.e. $\Delta^\alpha{}_\beta=0$, and the
only external current is the electromagnetic energy--momentum current
$\Sigma_\alpha$, given by 
\begin{equation}
\Sigma_\alpha = 2 a_0 \,\left( e_\alpha\rfloor V_{{\rm
      Max}}+(e_\alpha\rfloor F)\wedge H \right)\, .
\label{electro}
\end{equation}

The monopole type solutions are found in terms of isotropic coordinates 
$(t,r,\theta,\phi)$, which lead to the coframe \cite{mms98} 
\begin{equation}
\vartheta ^{\hat{0}} =\, \frac{1}{f}\, d\,t \,,\quad
\vartheta ^{\hat{1}} =\, f\, d\, r\, , \quad
\vartheta ^{\hat{2}} =\, f\,r\, d\,\theta\,,\quad
\vartheta ^{\hat{3}} =\, f\,r\, \sin\theta \, d\,\varphi
\label{frame2}\, ,
\end{equation}
with one unknown function $f=f(r,\theta)$. The coframe is assumed to be {\em
orthonormal}, that is, with the local Minkowski metric
$o_{\alpha\beta}:=\hbox{diag}(-1,1,1,1) =o^{\alpha\beta}$, yielding the 
spherically symmetric metric in {\em isotropic} form 
\begin{equation} 
ds^2=o_{\alpha\beta}\,\vartheta^\alpha\otimes\vartheta^\beta 
= - \frac{1}{f^2}\,dt^2+ f^2 \left[dr^2
+ r^2\left(d\theta^2+\sin^2\theta \,d\phi^2\right)\right] 
\label{gps}\, .
\end{equation}

As for the torsion and nonmetricity configurations, we concentrate on
the simplest non--trivial case with shear. The nonmetricity,
according to its irreducible decomposition, contains two covector pieces, 
namely $^{(4)}Q_{\alpha\beta}= Q\,g_{\alpha\beta}$, the dilation piece, and
\begin{equation}
  ^{(3)}Q_{\alpha\beta}={4\over
    9}\left(\vartheta_{(\alpha}e_{\beta)}\rfloor \Lambda - {1\over
      4}g_{\alpha\beta}\Lambda\right)\,,\qquad \hbox{with}\qquad
  \Lambda:= \vartheta^{\alpha}e^{\beta}\rfloor\!
  {\nearrow\!\!\!\!\!\!\!Q}_{\alpha\beta}\label{3q}\,,
\end{equation}
a proper shear piece \cite{nehe} with 
${\nearrow\!\!\!\!\!\!\!Q}_{\alpha\beta}:=Q_{\alpha\beta}-Q\,g_{\alpha\beta}$ 
the remaining traceless piece of the nonmetricity. Accordingly, our ansatz
for the nonmetricity reads
\begin{equation}
  Q_{\alpha\beta}=\, ^{(3)}Q_{\alpha\beta} +\,
  ^{(4)}Q_{\alpha\beta}\,.\label{QQ}
\end{equation}
The torsion, in addition to its tensor piece,
encompasses a vector and an axial vector piece. Let us choose only
the vector piece as non--vanishing:
\begin{equation}
T^{\alpha}={}^{(2)}T^{\alpha}={1\over 3}\,\vartheta^{\alpha}\wedge T\,,
\qquad \hbox{with}\qquad T:=e_{\alpha}\rfloor T^{\alpha}\,.\label{TT}
\end{equation}

Thus we are left with the three non--trivial one--forms $Q$, $\Lambda$,
and $T$, which should not distinguish a direction in space in the spherically 
symmetric case.
The following ansatz, for the above mentioned triplet of one--forms, turns 
out to be compatible with that 
condition,
\begin{equation}
Q=k_0 \left[ N_e u(r,\theta)\,\vartheta^{\hat{0}}+
N_g \frac{w(r,\theta)}{f\, r \sin\theta}\,\vartheta^{\hat{3}} \right]
=\frac{k_0}{k_1}\Lambda=\frac{k_0}{k_2}T
\label{genEug}\, ,
\end{equation}
with $N_e$ the gravito--electric charge and $N_g$ the gravito--magnetic one.
The electromagnetic potential $A$ appropiate for this configuration can be
expressed as follows \cite{helo,aas}
\begin{equation}
A= e_o\, u(r,\theta) \vartheta^{\hat{0}}
+ g_o\, \frac{w(r,\theta)}{f\,r \sin \theta}\vartheta^{\hat{3}}
\label{emch}\, ,
\end{equation}
where $e_o$ is the electric charge and $g_o$ the magnetic charge.
\section{Electrically and magnetically charged strong gravito--magnetic 
monopoles}

In order to solve the equations arising from the MAG Lagrangian (\ref{lobo}), 
we removed all other curvature square pieces by setting 
$w_{I}=0$, $z_{1}=z_{2}=z_{3}=z_{5}=0$, $b_5=0$ and $\lambda = 0$ in the  
Lagrangian
keeping the square of the segmental curvature in the Lagrangian
\cite{heh96,heh961}.

We substitute the local metric $o_{\alpha\beta}$, the coframe
(\ref{frame2}), the nonmetricity and torsion (\ref{genEug}) into the
field equations (\ref{first}), (\ref{second}) of the Lagrangian
(\ref{lobo}). 
Provided the (rather weak) constraint 
\begin{eqnarray}
32&a_0^2& b_4 - 4a_0 a_2 b_4+64 a_0 b_3 b_4- 32 a_2 b_3 b_4 + 48 a_0 b_4 c_3
+ 24 b_4 c_3^2 + 24 b_3 c_4^2 \nonumber \\
&+& 12 a_0 a_2 b_3 +48 a_0 b_3 c_4 - 9a_0 c_3^2 + 18 a_0 c_3 c_4 + 3a_0 c_4^2
+6a_0^2 a_2 + 24 a_0^2 c_4 =0
\label{const} \, ,
\end{eqnarray}
on the coupling constants is fulfilled, then we find an exact solution with
\begin{equation}
f=1 \pm q u(r) \, , \qquad u(r)=1 - \frac{1}{r}\, , \quad 
w= \left(1-\cos \theta \right) 
\label{f3}\, .
\end{equation}
Here $q$ is an  arbitrary 
{\em integration constant}, and the coefficients $k_{0}, k_{1}, k_{2}$ in
the ansatz (\ref{genEug}) are determined by the 
dimensionless coupling constants of the Lagrangian:
\begin{eqnarray}
k_0 &=& \left({a_2\over 2}-a_0\right)(8b_3 + a_0) - 3(c_3 + a_0 )^2\,,
\label{k0}\\
k_1 &=& -9\left[ a_0\left({a_2\over 2} - a_0\right) + 
(c_3 + a_0 )(c_4 + a_0 )\right]\,,
\label{k1}\\
k_2 &=& {3\over 2} \left[ 3a_0 (c_3 + a_0 ) + (
8b_3 + a_0)(c_4 + a_0 )\right]
\label{k2}\, .
\end{eqnarray}
Then the constraint (\ref{const}) can be put into the following more compact 
form
\begin{equation}
  b_4=\frac{a_0k+2c_4k_2}{8k_0}\,,\qquad\hbox{with}\qquad k:=
  3k_0-k_1+2k_2
\label{b4}\, ,
\end{equation}
implying the following relation for $z_4$
\begin{equation}
q^2= \kappa \left(e_o^2 + g_o^2+  z_4 k_0^2  \frac{ N_e^2 + N_g^2}{2a_0}\right)
\label{z4}\, .
\end{equation}
The factor $k$ is directly related to the antisymmetric piece of the 
Ricci 1--form \cite{heh96}.
The electromagnetic potential can now be written as
\begin{equation}
A= e_o\, \left(1 - \frac{1}{r}\right) \vartheta^{\hat{0}}
+ g_o\, \frac{1-\cos \theta}{f\,r \sin \theta}
\vartheta^{\hat{3}}
\label{emch0}\, .
\end{equation}

Collecting our results, the nonmetricity and the torsion read
as follows: 
\begin{equation}
Q^{\alpha\beta}=\left[k_0 \,o^{\alpha\beta}+\frac{4}{9}\,
k_1 \,\left(\vartheta^{(\alpha}e^{\beta )}\rfloor-\frac{1}{4}\,
o^{\alpha\beta}\right)\right]
\left[N_e\, \left(1 - \frac{1}{r}\right)\, \vartheta^{\hat{0}} 
+ N_g\,\frac{1-\cos \theta}{f\,\sin\theta}
\vartheta^{\hat{3}}\right]
\label{nichtmetrizitaet}\,,
\end{equation}
\begin{equation}
T^\alpha= \frac{ \, k_2}{3}\vartheta^\alpha\wedge\,
\left[N_e\, \left(1 - \frac{1}{r}\right)\,\vartheta^{\hat{0}} 
+ N_g\,\frac{1-\cos \theta}{f\,\sin\theta}\vartheta^{\hat{3}}\right]
\label{tosion}\, .
\end{equation}

The solutions presented in the section correspond to gravito--electric, 
gravito--magnetic, electric and magnetic charged monoples.
\section{Monopole solution in Schwarzschild--like coordinates}

As it is well known, the Schwarzschild coordinates are not always the more 
suitable coordinates for finding exact solutions. Nevertheless,
we proceed in this section to write the monopole solutions of
Sec. IV in Schwarzschild coordinates in order to compare 
them with the Reissner--Nordstr\"om solution of Sec. VII. 

Therefore, let us rewrite the solution (\ref{f3}) in   
Schwarzschild coordinates $(\tau,\rho,\theta,\phi)$. In Eq. (\ref{frame2}),
we consider the explicit form of $\vartheta^{\hat 2}$ and 
$\vartheta^{\hat 3}$. Accordingly, we have to require \cite{hema99}
\begin{equation}
\rho= f\, r= \left[1\pm q u(r) \right]\, r =  
\left(1\pm q\right)\,r \mp q\, , \quad 
dr=\frac{d\rho}{1 \pm q}\, ; 
\quad \tau=\frac{t}{1\pm q} \, .
\end{equation}
Thus
\begin{equation}
f= \frac{1\pm q}{g}\, ; \qquad 
u(\rho)= 1- \frac{1}{(\rho\pm q)/(1\mp q)}=
1- \frac{1\mp q}{\rho\pm q}= \frac{\rho-1}{\rho \pm q}=\frac{1 - 1/\rho}{g}\, ,
\end{equation}
where the function $g(\rho)$ reads
\begin{equation}
g(\rho)= 1\pm \frac{ q}{\rho}
\label{globo}\, .
\end{equation}
Therefore, in Schwarschild coordinates, the coframe turns out to be 
\begin{equation}
\vartheta ^{\hat{0}} =\, g(\rho) \, d\, \tau \,,\quad
\vartheta ^{\hat{1}} =\, \frac{1}{g(\rho)} d\, \rho\, , \quad
\vartheta ^{\hat{2}} =\, \rho\, d\,\theta\,,\quad
\vartheta ^{\hat{3}} =\, \rho\, \sin\theta \, d\,\phi
\label{frame4}\, .
\end{equation}
It is {\em  orthonormal} and the metric reads
\begin{equation} 
ds^2=o_{\alpha\beta}\,\vartheta^\alpha\otimes\vartheta^\beta 
= - g^2\,d \tau^2
+ \frac{1}{g^2} d\rho^2
+\rho^2\left(d\theta^2+\sin^2\theta \,d\phi^2\right) 
\label{gps2}\, .
\end{equation}
   
The electromagnetic potential (\ref{emch}) for this solution 
can be written as
\begin{equation}
A= e_o\, \frac{1- 1/\rho}{g} \vartheta^{\hat{0}}
+ g_o\,\frac{1-\cos\theta}{\rho \sin \theta}
\vartheta^{\hat{3}}
\label{emch2}\, .
\end{equation}
The relation
\begin{equation}
q^2=\kappa \left( e_o^2 + g_o^2+ z_4 k_0^2  \frac{ N_e^2 + N_g^2}{2a_0} \right)
\label{z46}\, ,
\end{equation}
remains the same.

The nonmetricity and the torsion read
as follows:
\begin{equation}
Q^{\alpha\beta}=\left[k_0 \,o^{\alpha\beta}+\frac{4}{9}\,
k_1 \,\left(\vartheta^{(\alpha}e^{\beta )}\rfloor-\frac{1}{4}\,
o^{\alpha\beta}\right)\right]
\left[ N_e\, \frac{1- 1/\rho}{g}\vartheta^{\hat{0}} 
+ N_g\,\frac{1- \cos\theta}{\rho \sin\theta}
\vartheta^{\hat{3}}\right]
\label{nichtmetrizitaet4}\,,
\end{equation}
\begin{equation}
T^\alpha= \frac{k_2}{3}\vartheta^\alpha\wedge\,
\left[ N_e\, \frac{1- 1/\rho}{g}\vartheta^{\hat{0}} 
+ N_g\,\frac{1-\cos \theta}{\rho \sin\theta}\vartheta^{\hat{3}}\right]
\label{torsion6}\, .
\end{equation}

Now we are in position to present the generalized Reissner--Nordstr\"om 
solution in MAG and to compare it with our monopole solutions.
\section{Generalized Reissner--Nordstr\"om solution in MAG}

In order to obtain the generalized Reissner--Nordstr\"om solution in MAG
we choose a coframe of the Schwarzschild type \cite{gama}
\begin{equation}
\vartheta^0= f(r) dt, \qquad \vartheta^1= \frac{1}{f(r)} dr, \qquad
\vartheta^2= r d\theta ,  \qquad 
\vartheta^3= r \sin \theta d\phi \, ,
\label{coframe}
\end{equation}
which leads to the specific spherically symmetric metric
\begin{equation}
ds^2= - f^2\,d t^2
+ \frac{1}{f^2} d r^2
+r^2\left(d\theta^2+\sin^2\theta \,d\phi^2\right) 
\label{metric}\, ,
\end{equation}
where the function $f(r)$ is unknown. The ansatz for the triplet
of one--forms is now given by 
\begin{equation}
Q=k_0 \left[ N_e u(r,\theta)\,\vartheta^{\hat{0}}+
N_g \frac{w(r,\theta)}{r \sin\theta}\,\vartheta^{\hat{3}} \right]
=\frac{k_0}{k_1}\Lambda=\frac{k_0}{k_2}T
\label{genEug1}\, .
\end{equation}

The ansatz for the electromagnetic potential 
turns out to be \cite{helo,aas}
\begin{equation}
A= e_o\, u(r,\theta)\, \vartheta^{\hat 0} 
+ g_o\, \frac{w(r,\theta)}{r \sin \theta}\, \vartheta^{\hat 3}
\label{potential}\, ,
\end{equation}
where $\rm e_o$ is the electric charge and $\rm g_o$ the magnetic charge 
respectively.

For $\lambda \neq 0$ we obtain the following solution
for the structure functions
\begin{eqnarray}
f(r)&:=& {\rm \sqrt{1- \frac{2M}{r}+ \kappa \frac{e_o^2+g_o^2 
+(z_4/2a_0) k_0^2\left(
N_e^2+N_g^2\right)}{r^2}
-\frac{\lambda }{3a_0}r^2 }  }\label{f2} \, ,\\
u(r)&:=& 1/f\, r\label{u2} \, ,\\
w(\theta)&:=&{\rm \left(1- cos \theta  \right)} . \label{w2}
\end{eqnarray}
Thus, the electromagnetic potential has 
the following form
\begin{eqnarray}
A&=& e_o\, \frac{1}{f\,r} \, \vartheta^{\hat 0}
+ g_o\, \frac{1-\cos\theta}{r \sin \theta}\, \vartheta^{\hat 3}
\nonumber \\
&=& e_o\, \frac{1}{r} \, dt
+ g_o\, \left(1-\cos\theta \right) d \phi
\label{potential2}\, .
\end{eqnarray}
Moreover, the nonmetricity and the torsion can therefore be expressed as
\begin{equation}
Q^{\alpha\beta}=\left[k_0 \,o^{\alpha\beta}+\frac{4}{9}\,
k_1 \,\left(\vartheta^{(\alpha}e^{\beta )}\rfloor-\frac{1}{4}\,
o^{\alpha\beta}\right)\right]
\left[ N_e\, \frac{1}{f\, r}\vartheta^{\hat{0}} 
+ N_g\,\frac{1-\cos \theta}{r \sin\theta}
\vartheta^{\hat{3}}\right]
\label{nichtmetrizitaet5}\,,
\end{equation}
\begin{equation}
T^\alpha= \frac{k_2}{3}\vartheta^\alpha\wedge\,
\left[ N_e\, \frac{1}{f\, r}\vartheta^{\hat{0}} 
+ N_g\,\frac{1-\cos \theta}{r \sin\theta}\vartheta^{\hat{3}}\right]
\label{torsion7}\, .
\end{equation}
Let us summarize the properties of the MAG--Maxwell solution presented
here. The function $f$, which fixes the orthonormal coframe (\ref{coframe}),
has six contributions, see (\ref{f2}). The terms containing the mass parameter 
$M$, the cosmological constant $\lambda$, and the electric charge $e_o$ 
correspond exactly to the general relativistic Reissner--Nordstr\"om 
solution with cosmological constant. The term with the gravito--electric
dilation charge $N_e$ has a similar structure as the previous term with the 
electric charge $e_o$.
Additionally, we have in (\ref{f2}) a magnetic charge $g_o$ term 
and a gravito--magnetic charge $N_g$ term. Therefore, our solution is
a generalizad Reissner--Nordstr\"om solution. 
It is clear that all relevant geometric 
objects, i.e. coframe, connection, torsion, curvature and nonmetricity
{\em feel} via the function $f$ the presence of the electromagnetic field.
However, otherwise the Maxwell field is disconnected from nonmetricity
and torsion.
  
It is interesting to note that if we set $g_o=0$ and $N_g=0$ in the 
generalized Reissner--Nordstr\"om solution presented above, see Eqs.
Eqs. (\ref{f2})--(\ref{torsion7}), the solution of Puntigam et al. 
\cite{PLH97} is obtained. Moreover, by setting $N_g=0$, $A=0$ the solution 
of Obukov et al. \cite{heh96} is recovered. Besides, for $M=e_o$, i.e.
mass equal to charge, and cosmological constant $\lambda=0$, 
it reduces to the monopole solution 
Eqs. (\ref{globo}), and (\ref{emch2})--(\ref{torsion6}). 
Additionally, by setting $N_g$ and $A=0$ in the charged monopole solution 
one recovers the vacuum gravito--electric charged monopole solution of 
Mac\'{\i}as et al. \cite{mms98}, and  the first term of the 
multipole solutions of Socorro et al. \cite{slmm}. 
\section{Further multipole solutions in MAG}

Further solutions of the same type and in the framework of the same
monopole configuration of Sec. III can be obtained by taking
\begin{equation}
f=1 \pm q u(r,\theta)\, , \quad u(r,\theta)= \frac{\cos\theta}{r^2}\, , 
\quad
w(r,\theta)= \, \frac{\sin^2 \theta}{r}
\label{mono2}\, ,
\end{equation}
with an electromagnetic potential $A$ given by
\begin{equation}
A= e_o \,\frac{\cos\theta}{r^2} \vartheta^{\hat{0}}  
+ g_o\, \frac{\sin \theta}{r^2\, f}
\vartheta^{\hat{3}}
\label{emch1}\, .
\end{equation}
The nonmetricity and the torsion can now be written as
\begin{equation}
Q^{\alpha\beta}=\frac{1}{r}\,\left[k_0 \,o^{\alpha\beta}+\frac{4}{9}\,
k_1 \,\left(\vartheta^{(\alpha}e^{\beta )}\rfloor-\frac{1}{4}\,
o^{\alpha\beta}\right)\right]
\left[ N_e \frac{\cos\theta}{r}\vartheta^{\hat{0}} 
+ N_g \frac{\sin \theta}{r\,f} \vartheta^{\hat{3}}\right]
\label{nichtmetrizitaet1}\,,
\end{equation}
\begin{equation}
T^\alpha= \frac{k_2}{3\,r}\vartheta^\alpha\wedge\,
\left[ N_e \,\frac{\cos \theta}{r}\vartheta^{\hat{0}} 
+ N_g \frac{\sin \theta}{r\, f} \, \vartheta^{\hat{3}}\right]
\label{tosion1}\, .
\end{equation}

The relation 
\begin{equation}
q^2=  \kappa \left( e_o^2 + g_o^2+  z_4 k_0^2  \frac{ N_e^2 + N_g^2}{2a_0} 
\right)
\label{z41}\, ,
\end{equation}
for $z_4$ still stands.
These solutions are also endowed with gravito--electric $N_e$ and 
gravito--magnetic $N_g$ charges as well as with electric $e_o$ and 
magnetic $g_o$ charges.

Although the solutions presented in this section are {\em not} particular 
cases of the avobe presented generalized Reissner--Nordstr\"om solution,
they are also interesting solutions which contribute to enhance our 
physical insight in the MAG theories.

All these solutions were checked with Reduce \cite{REDUCE} with its Excalc 
package \cite{EXCALC} for treating exterior differential forms\cite{Stauffer} and 
the Reduce--based GRG computer algebra system \cite{GRG,saf98}.
\section{Charge assignmemt of the solutions}

With propagating nonmetricity $Q_{\alpha\beta}$ two types of charge are
expected to arise: {\em One dilation charge} related by the Noether procedure 
to 
the trace $Q:=Q_\gamma{}^\gamma/4$ of the nonmetricity, called the Weyl 
covector $Q=Q_i dx^i$. It is the connection associated with gauging $R^+$
instead of $U(1)$ in the case of the Maxwell potential $A=A_i dx^i$. 
{\em Nine} types of {\em shear charge} are related to the remaining traceless 
piece
${\nearrow\!\!\!\!\!\!\!Q}_{\alpha\beta}$ 
of the nonmetricity. Under the local Lorentz group, the nonmetricity can be 
decomposed into four irreducible pieces $^{(I)}Q_{\alpha\beta}$, with 
$I=1,2,3,4$. 
The Weyl covector is linked to $^{(4)}Q_{\alpha\beta}=Q\,g_{\alpha\beta}$. 
Therefore we should find $4+4+1$ shear charges and 1 dilation charge.

Besides mass $M$, monopole charges $e_o$ and $g_o$, our solution carries 
dilation, shear, and spin charges, each of them of the covectorial type. 
We have the following assignments:
\begin{eqnarray}\label{chargem}
{M} \>\; &\longrightarrow& \hbox{mass of Schwarzschild type}\,,\\
{e_o} \>\; &\longrightarrow& \hbox{electric charge}\,,\\ 
{g_o} \>\; &\longrightarrow& \hbox{magnetic charge}\,,\\ 
{N_e} \>\; &\longrightarrow& \hbox{gravito--electric charge}\,,\\ 
{N_g} \>\; &\longrightarrow& \hbox{gravito--magnetic charge}\,,\\ 
\label{chargek0}
k_0 N_e, \, k_0 N_g &\longrightarrow& \hbox{dilation (`Weyl') charges of type 
$^{(4)}Q^{\alpha\beta}$}\,,\\ \label{chargek1}
k_1 N_e, \, k_1 N_g &\longrightarrow& \hbox{shear charges of type  
$^{(3)}Q^{\alpha\beta}$}\,,\\ \label{chargek2}
k_2 N_e, \, k_2 N_g &\longrightarrow& \hbox{spin charges of type 
$^{(2)}T^\alpha$}\,.
\end{eqnarray}      
In principle, however, these ``charge" assignments need to be {\em justified} 
by integrating locally conserved Noether currents in MAG, such as Eq. 
(5.7.4) of Ref. \cite{PR}.

\section{Discussion}

The constants $a_1\,, a_3\,; b_1\,,b_2\, , c_2$ do not occur in our
Reissner--Nordstr\"om solution (\ref{f2})--(\ref{torsion7}), nor in the 
constraint (\ref{b4}), so they are irrelevant and we can put them to zero,
Then the Lagrangian (\ref{Ltot}) reduces to
\begin{eqnarray}
L&=&  V_{MAG} + V_{Max} \nonumber \\
&=& \frac{1}{2\kappa}\,\left[-a_0\,R^{\alpha\beta}\wedge
\eta_{\alpha\beta}-2\lambda\,\eta+
a_{2}\,T^\alpha\wedge{}^*\!\,^{(2)}T_\alpha\right.
\nonumber\\&+&\left.
2\left(c_{3}\,^{(3)}Q_{\alpha\beta}+c_{4}\,^{(4)}
Q_{\alpha\beta}\right)\wedge\vartheta^\alpha
\wedge{}^*\! T^\beta\right.\nonumber\\& +& 
\left.Q_{\alpha\beta}\wedge{}^*\!\left(b_{3}\,^{(3)}
Q^{\alpha\beta}+b_{4}\,^{(4)}Q^{\alpha\beta}\right)\right]
\nonumber\\&-& \frac{z_{4}}{2}\,R^{\alpha\beta} 
\wedge{}^*\!\,^{(4)}Z_{\alpha\beta} - \frac{1}{2}\, F \wedge {}^*\, F
\,,\label{nondeg1}
\end{eqnarray}
where $b_4$ is determined by (\ref{b4}). 
Consequently our solutions (\ref{f2})--(\ref{torsion7}), with 
(\ref{k0})--(\ref{z4}), solve also the field equations belonging to
(\ref{nondeg1}). However, for the monopole and
multipole solutions (\ref{f3})--(\ref{tosion}) 
and (\ref{mono2})--(\ref{z41}), respectively, we set additionally 
$\lambda=0$.
It is interesting to note that, in
(\ref{nondeg1}) nonmetricity and torsion enter explicitly only with
those pieces which are admitted according to (\ref{chargek0}),
(\ref{chargek1}), (\ref{chargek2}).

In view of \cite{tw,oveh97}, it is clear that one may start with any
solution of the Einstein--Maxwell equations, then one replaces, after
imposing a suitable constraint on the coupling constants, the
electric and/or magnetic charge by strong gravito charges therby arriving
at the post--Riemannian triplet. Now, one chooses an electromagnetic 
potential one--form $A$
proportional to the triplet one--form. Thus the structure of the
energy--momentum of the triplet one--form and of the one--form $A$ is
the same one. Moreover, both currents differ only by a constant. 
Explicit solutions convey a better undestanding of the structures
involved.

In order to obtain solutions which excite more general post--Riemannian 
structures, it is desirable to go beyond the triplet ansatz 
(\ref{genEug}), which requires a generalization of the Lagrangian 
(\ref{nondeg1}).

The theories of modern physics generally involve a mathematical model,
defined by a certain set of differential equations, and supplemented 
by a set of rules for translating the mathematical results into meaninful
statements about the physical world. In the case of gravity theories,
because they deal with the most universal of the physical interactions,
one has an additional class of problems concerning the influence of the
gravitational field on other fields and matter. These are often studied
by working within a fixed gravitational field, usually an exact 
solution \cite{kr80}.
In this context our generalized Reissner--Nordstr\"om solution and the
monopole solutions contained on it contribute to enhance our understanding
of some of these questions in the framework of MAG theories, in particular
the ones concerned with the electrovacuum sector of them \cite{nehe,aas}.

\acknowledgments

We thank Friedrich W. Hehl for useful discussions and literature hints.
This research was supported by CONACYT Grants 28339E and 3898P-E9608,
and by the joint German--Mexican project DLR--Conacyt 
E130--2924 and MXI 009/98 INF .

\end{document}